\documentclass[reprint,twocolumn,pre,showpacs,amsmath,amssymb,aps]{revtex4-1}

\usepackage{natbib}	
\usepackage{graphicx,color,rotating}
\usepackage[latin1]{inputenc}
\usepackage{textcomp}
\usepackage{dcolumn}
\usepackage{bm}     
\usepackage{upgreek}
\usepackage[latin1]{inputenc}

\usepackage{array}
\newcolumntype{L}[1]{>{\raggedright\let\newline\\\arraybackslash\hspace{0pt}}m{#1}}
\newcolumntype{C}[1]{>{\centering\let\newline\\\arraybackslash\hspace{0pt}}m{#1}}
\newcolumntype{R}[1]{>{\raggedleft\let\newline\\\arraybackslash\hspace{0pt}}m{#1}}
\newcommand\Tstrut{\rule{0pt}{4.0ex}}         

\newcommand{\notop}{{{}_{}}}

\renewcommand{\vec}[1]{\bm{#1}}
\newcommand{\ee}{\mathrm{e}}
\newcommand{\ii}{\mathrm{i}}
\newcommand{\dm}{\mathrm{d}}
\newcommand{\dd}{\mathrm{d}}
\newcommand{\avr}[1]{\big\langle #1 \big\rangle}

\DeclareMathOperator{\re}{Re}

\newcommand{\iot}{{\ii\omega t}}

\newcommand{\ve}{\varepsilon}
\newcommand{\veO}{\ve^\notop_0}
\newcommand{\veI}{\ve^\notop_1}

\newcommand{\pp}{\partial^{{}}}
\newcommand{\ppsqr}{\partial^{\,2_{}}}

\newcommand{\nablabf}{\boldsymbol{\nabla}}

\newcommand{\Lapl}{\nabla^2}

\newcommand{\ppFixed}[3]{\bigg(\frac{\pp #1}{\pp #2}\bigg)^{{}}_{\! #3}}

\newcommand{\ie}{\textit{i.e.}}
\newcommand{\eg}{\textit{e.g.}}

\newcommand{\etal}{\textit{et~al.}}

\newcommand{\fracsmall}[2]{\mbox{$\frac{#1}{#2}$}}

\newcommand{\fff}{\vec{f}}

\newcommand{\nnn}{\vec{n}}
\newcommand{\nnnn}{\vec{n}^\notop}

\newcommand{\Pin}{P^\notop_\mathrm{in}}
\newcommand{\Pout}{P^\notop_\mathrm{out}}

\newcommand{\rrr}{\vec{r}}

\newcommand{\vvv}{\vec{v}}

\newcommand{\vstr}{v^{{}}_\mathrm{str}}

\newcommand{\vstrR}{v^\mathrm{R}_\mathrm{str}}

\newcommand{\vplu}{v_\mathrm{plug}}
\newcommand{\vpoi}{v_\mathrm{poi}}

\newcommand{\zerovec}{\boldsymbol{0}}

\newcommand{\Cp}{c^\notop_p}
\newcommand{\cp}{c^\notop_p}

\newcommand{\cV}{c^\notop_V}

\newcommand{\Dth}{D^\notop_\mathrm{th}}

\newcommand{\Eac}{E^{{}}_\mathrm{ac}}

\newcommand{\kth}{k^\mathrm{th}}
\newcommand{\kthO}{k^\mathrm{th}_0}
\newcommand{\kthI}{k^\mathrm{th}_1}

\newcommand{\kapT}{\kappa^\notop_T}

\newcommand{\kaps}{\kappa^\notop_s}

\newcommand{\Ps}{P^{{}}_\mathrm{s}}

\newcommand{\vbc}{v^\notop_\mathrm{bc}}

\newcommand{\alfP}{{\alpha^\notop_p}}

\newcommand{\delt}{\delta^\notop_\mathrm{t}}

\newcommand{\dels}{\delta^\notop_\mathrm{s}}
\newcommand{\delssqr}{\delta^2_\mathrm{s}}

\newcommand{\etaB}{\eta^\mathrm{b}}
\newcommand{\etaBO}{\eta^\mathrm{b}_0}
\newcommand{\etaBI}{\eta^\mathrm{b}_1}
\newcommand{\etab}{\eta^\mathrm{b}}

\newcommand{\etaO}{\eta^{{}}_0}
\newcommand{\etaI}{\eta^{{}}_1}

\newcommand{\dbd}{d_\mathrm{bd}}
\newcommand{\dbk}{d_\mathrm{bk}}
\newcommand{\cs}{c^{{}}_s}

\newcommand{\fO}{f^{{}}_0}

\newcommand{\fres}{f^{{}}_\mathrm{res}}

\newcommand{\gO}{g^{{}}_0}
\newcommand{\gI}{g^{{}}_1}
\newcommand{\gII}{g^{{}}_2}

\newcommand{\pO}{p^{{}}_0}

\newcommand{\pI}{p^{{}}_1}
\newcommand{\pIsqr}{p^{\,2_{}}_1}
\newcommand{\pIa}{p^\mathrm{a}_1}

\newcommand{\pII}{p^{{}}_2}

\newcommand{\sI}{s^{{}}_1}

\newcommand{\taubfI}{\bm{\tau}^{{}}_1}
\newcommand{\taubfII}{\bm{\tau}^{{}}_2}

\newcommand{\TO}{T^{{}}_0}
\newcommand{\TI}{T^{{}}_1}
\newcommand{\TII}{T^{{}}_2}
\newcommand{\TIIa}{T^\mathrm{a}_2}

\newcommand{\vO}{v^{{}}_0}
\newcommand{\vvvO}{\vvv^{{}}_0}

\newcommand{\vIsqr}{v^{\,2_{}}_1}

\newcommand{\vvvI}{\vvv^{{}}_1}

\newcommand{\vyI}{v^{{}}_{y1}}

\newcommand{\vyIa}{v^\mathrm{a}_{y1}}
\newcommand{\vvvIbc}{\vec{v}^\mathrm{bc}_{1}}
\newcommand{\vyIbc}{v^\mathrm{bc}_{y1}}
\newcommand{\vvvIres}{\vec{v}^\mathrm{res}_{1}}

\newcommand{\vyII}{v^{{}}_{y2}}

\newcommand{\vvvII}{\vvv^{{}}_2}

\newcommand{\rhoO}{\rho^\notop_0}
\newcommand{\rhoI}{\rho^\notop_1}

\newcommand{\SICel}{^\circ\!\textrm{C}}

\newcommand{\SIum}{\upmu\textrm{m}}

\newcommand{\SIK}{\textrm{K}}

\newcommand{\SImum}{\textrm{\textmu{}m}}

\newcommand{\SIMPa}{\textrm{MPa}}

\newcommand{\beq}[1]{\begin{equation} \eqlab{#1}}
\newcommand{\eeq}{\end{equation}}
\newcommand{\bsub}{\begin{subequations}}
\newcommand{\esub}{\end{subequations}}
\def\bal#1\eal{\begin{align}#1\end{align}}
\def\bsubal#1\esubal{\bsub \begin{align}#1\end{align} \esub}
\newcommand{\nn}{\nonumber}
\newcommand{\eqlab}[1]{\label{eq:#1}}
\renewcommand{\eqref}[1]{Eq.~(\ref{eq:#1})}
\newcommand{\eqsref}[2]{Eqs.~(\ref{eq:#1}) and~(\ref{eq:#2})}

\newcommand{\figref}[1]{Fig.~\ref{fig:#1}}

\newcommand{\figlab}[1]{\label{fig:#1}}

\newcommand{\secref}[1]{Section~\ref{sec:#1}}

\newcommand{\seclab}[1]{\label{sec:#1}}
\newcommand{\tabref}[1]{Table~\ref{tab:#1}}

\newcommand{\tablab}[1]{\label{tab:#1}}

\begin{document}

\title{A numerical study of thermoviscous effects in ultrasound--induced \\acoustic streaming in microchannels}

\author{Peter Barkholt Muller}
\email{peter.b.muller@fysik.dtu.dk}
\affiliation{Department of Physics, Technical University of Denmark, DTU Physics Building 309, DK-2800 Kongens Lyngby, Denmark}

\author{Henrik Bruus}
\email{bruus@fysik.dtu.dk}
\affiliation{Department of Physics, Technical University of Denmark, DTU Physics Building 309, DK-2800 Kongens Lyngby, Denmark}


\begin{abstract}\vspace*{-5mm}
\centerline{\small (Submitted to Phys.\ Rev.\ E, 21 August 2014)}

\vspace*{5mm}
We present a numerical study of thermoviscous effects on the acoustic streaming flow generated by an ultrasound standing-wave resonance in a long straight microfluidic channel containing a Newtonian fluid. These effects enter primarily through the temperature and density dependence of the fluid viscosity. The resulting magnitude of the streaming flow is calculated and characterized numerically, and we find that even for thin acoustic boundary layers, the channel height affects the magnitude of the streaming flow. For the special case of a sufficiently large channel height we have successfully validated our numerics with analytical results from 2011 by Rednikov and Sadhal for a single planar wall. We analyze the time-averaged energy transport in the system and the time-averaged second-order temperature perturbation of the fluid. Finally, we have made three main changes in our previously published numerical scheme to improve the numerical performance: (i) The time-averaged products of first-order variables in the time-averaged second-order equations have been recast as flux densities instead of as body forces. (ii) The order of the finite element basis functions has been increased in an optimal manner. (iii) Based on the International Association for the Properties of Water and Steam (IAPWS 1995, 2008, and 2011), we provide accurate polynomial fits in temperature for all relevant thermodynamic and transport parameters of water in the temperature range from 10~$\SICel$ to 50~$\SICel$.
\end{abstract}



\maketitle


\section{Introduction}
\seclab{Intro}

Ultrasound acoustophoresis has been used to handle particles of a few micrometer to tens of micrometer in microfluidic channels \cite{Bruus2011c}, with applications in \eg~up-concentration of rare samples \cite{Nordin2012}, cell syncronization \cite{Thevoz2010}, cell trapping \cite{Ohlin2013}, cell patterning \cite{Shi2009a}, cell detachment \cite{Bussonniere2014}, cell separation \cite{Augustsson2012} and particle rotation \cite{Schwarz2013}. Control and processing of sub-micrometer bioparticles have many application in biomedicine and in environmental and food analysis, however acoustophoretic focusing of sub-micrometer particles by the primary radiation force is hindered by the drag force from the acoustic streaming flow of the suspending liquid. Consequently, there is a need for understanding the acoustic streaming and for developing tools for engineering acoustic streaming patterns that allow for acoustic handling of sub-micrometer particles.

The theory of acoustic streaming, driven by the time-averaged shear stress near rigid walls in the acoustic boundary layers of a standing wave, was originally described by Lord Rayleigh \cite{LordRayleigh1884}.  It has later been extended, among others, by Schlicting \cite{Schlichting1932}, Nyborg \cite{Nyborg1958}, Hamilton \cite{Hamilton2003,Hamilton2003b}, and Muller \etal\ \cite{Muller2013}. Recently, Rednikov and Sadhal \cite{Rednikov2011} have included the temperature dependence of the dynamic viscosity and shown that this can lead to a significant increase of the magnitude of the streaming velocity. In the present work we present a numerical study of this and related thermoviscous effects.

A major challenge in numerical modeling of acoustic streaming is the disparate length scales characterizing the bulk of the fluid and the acoustic boundary layer, the latter often being several orders of magnitude smaller than the former in relevant experiments. One way to handle this problem is to determine the first-order oscillatory acoustic field without resolving the acoustic boundary layers, and from this calculate an approximate expression for the time-averaged streaming velocity at the boundary, acting as a boundary condition for the steady bulk streaming \cite{Nyborg1953a, Lee1989}. This method has the advantage of being computationally less demanding, such as Lei \etal\ \cite{Lei2013, Lei2014} used it to model streaming flow in microfluidic channels in three dimensions and were able to qualitatively explain several experimental observations of streaming flow in microchannels and flat microfluidics chambers. Another method is the direct numerical solution of the full thermoviscous acoustic equations both in the bulk and in the thin boundary layers, demanding a fine spatial resolution close to rigid surfaces as developed by e.g.\ Muller \etal~\cite{Muller2012}. They obtained a quantitative description of the physics of the thermoviscous boundary layers and the acoustic resonance. The same model was later employed in a quantitative comparison between numerics, analytics, and experiments of microparticle acoustophoresis, demonstrating good agreement \cite{Muller2013}. In a more recent study, the numerical scheme was further used to demonstrate how simultaneous actuation of the two overlapping half-wavelength resonances of a nearly-square channel can generate a single vortex streaming flow that allows for focusing of sub-micrometer particles, an effect demonstrated experimentally by focusing 0.5-$\upmu$m-diameter particles and \textit{E. coli} bacteria \cite{Antfolk2014}.

In this paper we extend our numerical model for a rectangular microchannel \cite{Muller2012} to include the thermoviscous effects, which were treated analytically in the special case of a single planar infinite rigid wall by Rednikov and Sadhal \cite{Rednikov2011}. The extension is done by including the dependence on the oscillatory first-order temperature and density fields in the dynamic shear viscosity, previously taken to be constant. This has a significant influence on the shear stresses in the thermoviscous boundary layers responsible for generating the steady acoustic streaming. Furthermore, we study the steady temperature rise and energy current densities resulting from solving the time-averaged second-order energy transport equation. Finally, we improve the convergence properties of our previous numerical scheme \cite{Muller2012} by implementing the governing equations in a source-free flux formulation and optimizing the order of the basis functions of the finite element scheme.

\section{Basic theory}
\seclab{theory}
In this section, we derive the governing equations for the first- and second-order perturbations to the thermoviscous acoustic fields in a compressible Newtonian fluid. We only consider the acoustics in the fluid, and treat the surrounding walls as ideal hard walls. Our treatment is based on textbook thermodynamics \cite{Landau1980} and thermoviscous acoustics \cite{Pierce1991}, but in a source-free flux formulation suitable for our specific numerical implementation. As water is our model fluid of choice, we carefully implement the best available experimental data for the thermodynamic and transport parameters provided by the International Association for the Properties of Water and Steam (IAPWS).

\subsection{Thermodynamics}
\seclab{thermodynamics}

The independent thermodynamic variables of the compressible Newtonian fluid are taken to be the temperature $T$ and the pressure $p$ \cite{Landau1980}. The dependent variables are the mass density $\rho$, the internal energy $\ve$ per mass unit and the entropy $s$ per mass unit. The first law of thermodynamics is usually stated with $s$ and $\rho$ as the independent variables,
\bsub
 \beq{FirstLaw}
 \dm \ve = T\:\dm s - p\: \dm \bigg(\!\frac{1}{\rho} \bigg)
 = T\: \dm s + \frac{p}{\rho^2}\: \dm\rho.
 \eeq
By a standard Legendre transformation of $\ve$ to the Gibbs free energy $g$ per unit mass, $g = \ve -Ts + p\:\frac{1}{\rho}$, we obtain the first law with $T$ and $p$ as the independent variables,
 \beq{Gibbs}
 \dm g = -s\:\dm T +\frac{1}{\rho}\: \dm p.
 \eeq
 \esub

Due to their importance in thermoacoustics, we furthermore introduce the following three thermodynamics coefficients: the isobaric heat capacity $\Cp$ per unit mass, the isobaric thermal expansion coefficient $\alfP$, and the isothermal compressibility $\kapT$, as
 \bsub
 \eqlab{thermo_quant}
 \bal
 \eqlab{cPdef}
 \Cp  &= T \ppFixed{s}{T}{p},\\[2mm]
 \eqlab{alfPdef}
 \alfP &= -\frac{1}{\rho}\: \ppFixed{\rho}{T}{p},\\[2mm]
 \eqlab{kapTdef}
 \kapT &= \frac{1}{\rho}\: \ppFixed{\rho}{p}{T}.
 \eal
Moreover, as a standard step towards getting rid of explicit references to the entropy, we derive from \eqsref{Gibbs}{alfPdef} the following Maxwell relation,
 \beq{dSdpT}
 \ppFixed{s}{p}{T} = - \frac{\ppsqr g}{\pp p \pp T}
 = -\ppFixed{(\frac{1}{\rho})}{T}{p} = -\frac{1}{\rho}\: \alfP.
 \eeq
 \esub
Using Eqs.~(\ref{eq:cPdef})-(\ref{eq:dSdpT}), we express $\dm s$ and $\dm \rho$ in terms of $\dm T$ and $\dm p$
 \bsub
 \eqlab{thermorelations}
 \bal
 \eqlab{Tds}
 T\dm s &= \Cp\:\dm T  - \frac{\alfP T}{\rho}\:\dm p,\\[2mm]
 \eqlab{drho}
 \frac{1}{\rho}\:\dm \rho &= \kapT\: \dm p - \alfP\:\dm T,
 \eal
which combined with \eqref{FirstLaw} lead to $\dm \ve$ in terms of $\dm T$ and $\dm p$
 \beq{dEpsTP}
 \rho\:\dm\ve = \big(\Cp \rho - \alfP p\big)\:\dm T
 + \big(\kapT p -  \alfP T\big)\:\dm p\:.
 \eeq
 \esub
Using Eqs.~(\ref{eq:Tds})-(\ref{eq:dEpsTP}), small changes  $\dm s$, $\dm\rho$, and $\dm\ve$ in the dependent thermodynamic variables $s$, $\rho$, and $\ve$ away from equilibrium can thus be expressed in terms of changes in the independent thermodynamic variables $T$ and $p$. In our numerical analysis, the default unperturbed equilibrium state is the one at ambient temperature $\TO = 25.0~\SICel$ and pressure $\pO = 0.1013~\SIMPa$.

\subsection{Physical properties of water (IAPWS)}\seclab{IAPWS}
The theoretical treatment of thermoviscous acoustics, requires detailed knowledge of the dependence on temperature and density (or temperature and pressure) of the physical properties of the fluid of choice. In the present paper, we use the parameter values for water supplied by the International Association for the Properties of Water and Steam (IAPWS) in its thorough statistical treatment of large data sets provided by numerous experimental groups \cite{Wagner2002, Huber2009, Huber2012}.

The values of the thermodynamic properties are taken from the IAPWS Formulation 1995 \cite{Wagner2002}, the shear viscosity is taken from the IAPWS Formulation 2008 \cite{Huber2009}, the thermal conductivity is taken from the IAPWS Formulation 2011 \cite{Huber2012}, while the bulk viscosity is taken from Holmes, Parker, and Povey \cite{Holmes2011}, who extended the work by Dukhin and Goetz \cite{Dukhin2009}. The IAPWS data set spans a much wider range in temperature and ambient pressures than needed in our work, and it is somewhat complicated to handle. Consequently, to ease the access to the IAPWS data in our numerical implementation, we have carefully fitted the temperature dependence of all properties at atmospheric pressure by fifth-order polynomials in temperature in the range from 10~$\SICel$ to 50~$\SICel$ as described in detail in Appendix~\ref{sec:appendix}. In the specified range, the differences between our fits and the IAPWS data are negligible. In \tabref{parameter_values} we have listed the physical properties of water at ambient temperature and pressure.

The thermodynamic coefficients of \eqref{thermorelations} are by definition evaluated at the equilibrium state $T=\TO$ and $p = \pO$ leaving all acoustics perturbations to enter only in the small deviations, e.g.\ $\dm T = \TI + \TII$. On the other hand, the transport coefficients of the fluid depend on the acoustic perturbation. To avoid the ambiguity of the pressure $p$ as either the ambient pressure outside the fluid or the intrinsic pressure (cohesive energy) of the fluid, we use \eqref{drho} to change variable from pressure $p$ to density $\rho$ in our treatment of the IAPWS data. To first order in the acoustic perturbation, we thus write the dynamic shear viscosity $\eta$, the bulk (second) viscosity $\etaB$, and the thermal conductivity $\kth$ as
\bsub
\bal
 \eqlab{etaperturbation}
	\eta(T,\rho) &= \etaO(\TO,\rhoO) + \etaI(\TO,\TI,\rhoO,\rhoI), \\
	\etaI &= \bigg(\frac{\partial \eta}{\partial T}\bigg)^{{}}_{T=\TO} \TI +
				\bigg(\frac{\partial \eta}{\partial \rho}\bigg)^{{}}_{\rho=\rhoO} \rhoI,\\[2mm]
	\etaB(T,\rho) &= \etaBO(\TO,\rhoO) + \etaBI(\TO,\TI,\rhoO,\rhoI),\\
	\etaBI &= \bigg(\frac{\partial \etaB}{\partial T}\bigg)^{{}}_{T=\TO} \TI +
				\bigg(\frac{\partial \etaB}{\partial \rho}\bigg)^{{}}_{\rho=\rhoO} \rhoI,\\[2mm]
	\kth(T,\rho) &= \kthO(\TO,\rhoO) + \kthI(\TO,\TI,\rhoO,\rhoI),\\
	\kthI &= \bigg(\frac{\partial \kth}{\partial T}\bigg)^{{}}_{T=\TO} \TI +
				\bigg(\frac{\partial \kth}{\partial \rho}\bigg)^{{}}_{\rho=\rhoO} \rhoI.
\eal\eqlab{transportcoeff}
\esub

For the acoustic amplitudes used in this model, the maximum relative perturbations, such as $|\etaI|/\etaO$, due to the temperature dependence of the transport coefficients, are 0.33\%, 0.53\%, and 0.034\% for $\eta$, $\etaB$, and $\kth$, respectively, and the perturbations due to the density dependence are 0.37\% and 0.82\% for $\eta$ and $\kth$, respectively. We could not find any literature on the density dependence of $\etaB$ of water.


\begin{table}[!t]
\caption{\label{tab:parameter_values}
IAPWS parameter values for pure water at ambient temperature 25\:$^\circ$C and pressure $0.1013$~MPa. For references see Appendix \ref{sec:appendix}.}
\begin{tabular}{ L{3.08cm} c  r@{ $\times$ }l  c }
\hline \hline
Parameter & Symbol & \multicolumn{2}{c}{Value} & Unit \\ \hline
\multicolumn{5}{l}{\textbf{Thermodynamic parameters:}}  \Tstrut \\[0.5mm]
Mass density
 &
$\rho$
 &
$9.970$ & $10^{2}$
 &
kg\:m$^{-3}$
\\[1.0mm]
Heat capacity
 &
$c^{{}}_p$
 &
$4.181$ & $10^{3}$
 &
J\:kg$^{-1}$\:K$^{-1}$
\\[1.0mm]
Speed of sound
 &
$c^{{}}_s$
 &
$1.497$ & $10^{3}$
 &
m\:s$^{-1}$
\\[1.0mm]
Compressibility
 &
$\kappa^{{}}_T$
 &
$4.525$ & $10^{-10}$
 &
Pa$^{-1}$
\\[1.0mm]
Thermal expansion
 &
$\alpha^{{}}_p$
 &
$2.573$ & $10^{-4}$
 &
K$^{-1}$
\\[1.0mm]
Heat capacity ratio
 &
$\gamma$
 &
$1.011$ & $10^{0}$
 &

\\[1.0mm]
\multicolumn{5}{l}{\textbf{Transport parameters:}}   \\[0.5mm]
Shear viscosity
 &
$\eta$
 &
$8.900$ & $10^{-4}$
 &
Pa\:s
\\[1.0mm]
Bulk viscosity
 &
$\eta^\mathrm{b}$
 &
$2.485$ & $10^{-3}$
 &
Pa\:s
\\[1.0mm]
Thermal conductivity
 &
$k^\mathrm{th}$
 &
$6.065$ & $10^{-1}$
 &
W\:m$^{-1}$\:K$^{-1}$
\\[1.0mm]
\multicolumn{5}{l}{\textbf{Thermodynamic derivatives:}} \\[0.5mm]

 &
$\dfrac{1}{\eta} \dfrac{\partial \eta}{\partial T} $
 &
$-2.278$ & $10^{-2}$
 &
K$^{-1}$
\\[2.5mm]

 &
$\dfrac{1}{\eta} \dfrac{\partial \eta}{\partial \rho} $
 &
$-3.472$ & $10^{-4}$
 &
kg$^{-1}$\:m$^3$
\\[2.5mm]

 &
$\dfrac{1}{\eta^\mathrm{b}} \dfrac{\partial \eta^\mathrm{b}}{\partial T} $
 &
$-2.584$ & $10^{-2}$
 &
K$^{-1}$
\\[2.5mm]

 &
$\dfrac{1}{k^\mathrm{th}} \dfrac{\partial k^\mathrm{th}}{\partial T}$
 &
$2.697$ & $10^{-3}$
 &
K$^{-1}$
\\[2.5mm]

 &
$\dfrac{1}{k^\mathrm{th}} \dfrac{\partial k^\mathrm{th}}{\partial \rho}$
 &
$2.074$ & $10^{-3}$
 &
kg$^{-1}$\:m$^3$
\\[2.5mm]
\hline \hline
\end{tabular}
\end{table}


\subsection{Governing equations}
\seclab{GovEq}

Besides the above thermodynamic relations, the governing equations of thermoviscous acoustics requires the introduction of the velocity field $\vvv$ of the fluid as well as the stress tensor $\bm{\sigma}$, which is given as \cite{bruus2008}
 \bsubal
 \eqlab{sigmaDef}
 \bm{\sigma} &= -p\:\bm{1} + \bm{\tau}, \\
 \bm{\tau} &= \eta\bigg[\nablabf\vvv + (\nablabf \vvv)^\mathrm{T}\bigg]
 + \bigg[\etaB -\frac{2}{3}\eta\bigg](\nablabf\cdot\vvv)\:\bm{1}.
 \esubal
Here, $\bm{1}$ is the unit tensor and the superscript "T" indicates tensor transposition.

Mass conservation implies that the rate of change $\pp_t\rho$ of the density in a test volume with surface normal vector $\nnn$ is given by the influx (direction $-\nnn$) of the mass current density $\rho\vvv$. In differential form by Gauss's theorem it is
 \bsub
 \beq{contEq}
 \pp_t \rho = \nablabf\cdot\big[-\rho\vvv\big].
 \eeq

Similarly, momentum conservation implies that the rate of change $\pp_t(\rho\vvv)$ of the momentum density in the same test volume is given by the stress forces $\bm{\sigma}$ acting on the surface (with normal $\nnn$), and the influx (direction $-\nnn$) of the momentum current density $\rho\vvv\vvv$. In differential form, neglecting body forces $\fff$, this becomes
 \beq{momentumEq}
 \pp_t (\rho\vvv) = \nablabf\cdot\big[\bm{\tau} - p\:\bm{1} - \rho\vvv\vvv\big].
 \eeq

Finally, energy conservation implies that the rate of change $\pp_t\big(\rho\ve + \frac{1}{2}\rho v^2\big)$ of the energy density (internal plus kinetic), is given by the power of the stress forces $\vvv\cdot\bm{\sigma}$ on the surface (direction $\nnn$), and the influx (direction $-\nnnn$) of both heat conduction power $-\kth\nablabf T$ and energy current density $(\rho\ve + \fracsmall{1}{2}\rho v^2)\vvv$. In differential form, neglecting heat sources in the volume, this becomes
 \beq{energyEq}
 \pp_t \big(\rho\ve + \fracsmall{1}{2}\rho v^2\big) =
 \nablabf\cdot\big[\vvv\cdot\bm{\tau} - p\:\vvv + \kth\nablabf T - \rho(\ve+\fracsmall{1}{2}v^2)\vvv\big].
 \eeq
 \esub

\subsection{First-order equations of \\thermoviscous acoustics}
\seclab{FirstOrder}
The homogeneous, isotropic quiescent state (thermal equilibrium) is taken to be the zeroth-order state in the acoustic perturbation expansion. Following standard first-order perturbation theory, all fields $g$ are written in the form $g = \gO + \gI$, for which $\gO$ is the value of the zeroth-order state and $\gI$ is the acoustic perturbation which must be much smaller than $\gO$. We assume that the acoustic perturbations $\gI$ are oscillating harmonically with the angular frequency $\omega$ of the acoustic actuation,
 \beq{fHarm}
 \gI(\rrr,t) = \gI(\rrr)\:\ee^{-\ii\omega t}, \qquad \pp_t \gI = -\ii\omega\gI.
 \eeq
For the velocity, the value of the zeroth-order state is $\vvvO = \zerovec$, and thus $\vvv = \vvvI$. The zeroth-order terms solve the governing equations for the zeroth-order state and thus drops out of the equations. Keeping only first-order terms, we obtain the first-order equations.

The continuity equation~(\ref{eq:contEq}) becomes
 \bsub
 \beq{massEqA}
 \pp_t\rhoI = -\rhoO\nablabf\cdot\vvvI,
 \eeq
which, by using \eqref{drho} in the form
 \beq{rho1B}
 \rhoI = \rhoO\Big[\kapT\:\pI - \alfP\:\TI \Big],
 \eeq
is rewritten to
 \beq{massEq1}
 \alfP\:\pp_t\TI - \kapT\:\pp_t\pI = \nablabf\cdot\vvvI.
 \eeq
 \esub
The momentum equation~(\ref{eq:momentumEq}) likewise becomes
 \bsub
 \beq{momentumEq1}
 \rhoO\pp_t\vvvI = \nablabf\cdot\big[\taubfI - \pI\bm{1}\big].
 \eeq
where $\taubfI$ is given by
 \beq{tau1}
\taubfI = \etaO\bigg[\nablabf\vvvI + (\nablabf \vvvI)^\mathrm{T}\bigg]
 + \bigg[\etaBO -\frac{2}{3}\etaO\bigg](\nablabf\cdot\vvvI)\:\bm{1}.
 \eeq
 \esub
The energy equation~(\ref{eq:energyEq}) requires a little more work. To begin with, it can be written as
 \bsub
 \beq{energyEqA}
 \rhoO\pp_t\veI + \veO\pp_t\rhoI = \kthO\Lapl\TI -\pO\nablabf\cdot\vvvI -\veO\rhoO\nablabf\cdot\vvvI.
 \eeq
The two terms containing $\veO$ cancel out due to the continuity equation~(\ref{eq:massEqA}), and the term $\rhoO\pp_t\veI$ is rewritten using \eqref{FirstLaw}, whereby
 \beq{energyEqB}
 \rhoO\TO\pp_t s^{{}}_1 + \frac{\pO}{\rhoO}\:\pp_t\rhoI = \kthO\Lapl\TI -\pO\nablabf\cdot\vvvI.
 \eeq
The two terms containing $\pO$ cancel out due to the continuity equation~(\ref{eq:massEqA}), and the term $\rhoO\TO\pp_t s^{{}}_1 $ is rewritten using the time derivative of \eqref{Tds}. This leads to
 \beq{energyEq1}
 \rhoO\Cp\:\pp_t \TI  - \alfP \TO\:\pp_t \pI = \kthO\Lapl\TI.
 \eeq
 \esub
Equations (\ref{eq:massEq1}), (\ref{eq:momentumEq1}), and (\ref{eq:energyEq1}) are the resulting first-order thermoviscous equations for conservation of mass, momentum, and energy, respectively. In the frequency domain they become
 \bsubal
 -\ii\omega \alfP\:\TI + \ii \omega \kapT\:\pI &= \nablabf\cdot\vvvI,\\
 -\ii\omega\rhoO\: \vvvI &= \nablabf\cdot\big[\taubfI - \pI\bm{1}\big],\eqlab{NSI}\\
 -\ii\omega \rhoO\Cp\:\TI  + \ii\omega \alfP \TO\: \pI &= \kthO\Lapl\TI. \eqlab{EconsI}
 \esubal

From \eqsref{NSI}{EconsI}, neglecting the pressure terms, we can derive the length scales $\dels$ and $\delt$ for diffusion of momentum and heat, respectively,
 \bsub
 \begin{alignat}{3}
 \dels &= \sqrt{\frac{2\etaO}{\rhoO \omega}}    &&= \sqrt{\frac{2\nu}{\omega}}  &&= 0.38~\SImum,\eqlab{dels}\\
 \delt &= \sqrt{\frac{2\kthO}{\rhoO \cp\omega}} &&= \sqrt{\frac{2\Dth}{\omega}} &&= 0.15~\SImum,\eqlab{delt}
 \end{alignat}
 \esub
where the subscript 's' indicates shear stress, subscript 't' indicates thermal, and $\nu=\etaO/\rhoO$ and $\Dth = \kth/(\rhoO\Cp)$ are the momentum and thermal diffusivities with numerical values derived from the parameter values at ambient temperature and pressure listed in \tabref{parameter_values}.

\subsection{Second-order time-averaged equations\\ of thermoviscous acoustics}
\seclab{SecondOrder}

Moving on to second-order perturbation theory, writing the fields as $g = \gO + \gI + \gII$, we note that the second-order acoustic perturbation $\gII$ may contain both oscillating terms and a time-constant term. The time averaging over one oscillation period of a field $g(t)$ is denoted $\avr{g}$. We note that all full time derivatives averages to zero, $\avr{\pp_t g(t)} = 0$.

In the following, all pure second-order fields are taken to be time averaged and thus written plainly as $\gII$ without the angled brackets. With this notation the second-order time-averaged continuity equation~(\ref{eq:contEq}) becomes
 \beq{massEq2}
 \nablabf\cdot\big[\rhoO\vvvII + \avr{\rhoI\vvvI}\big] = 0,
 \eeq
while the momentum equation~(\ref{eq:momentumEq}) takes the form
 \bsub\eqlab{momentumEq2both}
 \beq{momentumEq2}
 \nablabf\cdot\big[\taubfII -\pII\bm{1} - \rhoO\avr{\vvvI\vvvI}\big] = \zerovec,
 \eeq
where $\taubfII$ is given by
\bal
 \eqlab{tau2}
 \taubfII &= \etaO\bigg[\nablabf\vvvII + (\nablabf \vvvII)^\mathrm{T}\bigg]
 + \bigg[\etaBO -\frac{2}{3}\etaO\bigg](\nablabf\cdot\vvvII)\:\bm{1} \nn\\
 					 & + \Bigg\langle\!\etaI\bigg[\nablabf\vvvI\! + (\nablabf \vvvI)^\mathrm{T}\bigg]\Bigg\rangle
 + \Bigg\langle\bigg[\etaBI \!-\!\frac{2}{3}\etaI\bigg](\nablabf\!\cdot\!\vvvI)\:\bm{1}\!\Bigg\rangle.
  \eal
  \esub
It is in the two last terms that the temperature and density dependence of the viscosities come into play through the perturbations $\etaI$ and $\etaBI$.

The energy equation~(\ref{eq:energyEq}) in its second-order time-averaged form is initially written as
 \bsub
 \bal
 \eqlab{energyEq2A}
  \nablabf\cdot\Big[&\avr{\vvvI\cdot\taubfI} + \kthO\nablabf\TII + \avr{\kthI\nablabf\TI} - \pO\:\vvvII
  \\ \nn
  &- \avr{\pI\vvvI} - \veO\rhoO\vvvII - \veO\avr{\rhoI\vvvI} - \rhoO\avr{\veI\vvvI}\Big] = 0.
 \eal
The two terms with $\veO$ cancel due to the continuity equation~(\ref{eq:massEq2}). Next, using \eqref{FirstLaw}, we obtain the expression $\rhoO\veI = \rhoO\TO\sI + (\pO/\rhoO)\rhoI$, which upon insertion into \eqref{energyEq2A} leads to
 \bal
 \eqlab{energyEq2B}
  \nablabf\cdot\Big[&\avr{\vvvI\cdot\taubfI} + \kthO\nablabf\TII + \avr{\kthI\nablabf\TI} - \pO\:\vvvII
  \\ \nn
  & - \avr{\pI\vvvI} - \rhoO\TO\avr{\sI\vvvI} - \frac{\pO}{\rhoO}\avr{\rhoI\vvvI}\Big] = 0.
 \eal
The two $\pO$-terms cancel by the continuity equation~(\ref{eq:massEq2}). Then, from \eqref{Tds} we find $\rhoO\TO\sI = \rhoO\Cp\:\TI - \alfP\TO\:\pI$, which by substitution into \eqref{energyEq2B} yields
 \bal
 \eqlab{energyEq2}
 \nablabf\cdot\Big[&\kthO\nablabf\TII + \avr{\kthI\nablabf\TI} + \avr{\vvvI\cdot\taubfI}
 \\ \nn
  &- (1\!-\!\alfP\TO)\avr{\pI\vvvI} -\rhoO\Cp\avr{\TI\vvvI}\Big] = 0.
 \eal
 \eqlab{energyEq2all}\esub
Equations (\ref{eq:massEq2}), (\ref{eq:momentumEq2}), and (\ref{eq:energyEq2}) are the resulting time-averaged second-order thermoviscous acoustic equations for conservation of mass,
momentum, and energy, respectively.

The time-averaged acoustic energy density $\Eac$ in the fluid is given by \cite{Pierce1991}
\beq{Eac}
\Eac = \frac{1}{2}\kaps \avr{\pIsqr} + \frac{1}{2}\rhoO \avr{\vIsqr},
\eeq
where $\kaps = \kapT /\gamma$ is the isentropic compressibility and $\gamma = \cp/\cV$ is the ratio of specific heat capacities.

For a product of two time-harmonic fields in the complex-valued representation \eqref{fHarm}, the time average can be calculated as
 \beq{TimeAvr}
 \avr{\rhoI(\rrr,t)\: \vvvI(\rrr,t)} = \frac{1}{2}\:\re\big[\rhoI(\rrr,0)\:\vvv^{*_{}}_1(\rrr,0)\big],
 \eeq
where the asterisk denote complex conjugation.

%
\begin{figure}[t]
\centering
\includegraphics[width=0.99\columnwidth]{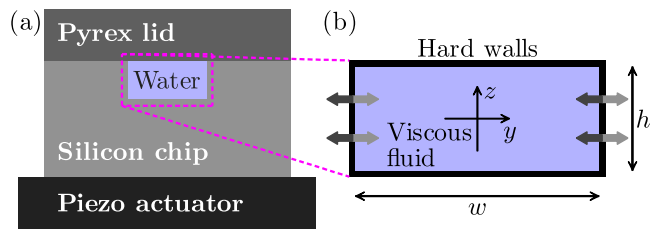}
\caption{\figlab{chip_sketch} (Color online) (a) Sketch of a physical acoustophoresis setup with a silicon microfluidic chip on top of a piezo actuator such as in \cite{Augustsson2011}. (b) Sketch of the model system for the numerical scheme with the viscous fluid domain surrounded by hard walls. The thick arrows indicate in-phase oscillating displacement of the left and right walls. Be default we set the width $w=380~\SImum$ and the height $h=160 \SImum$.}
\end{figure}%
\begin{figure*}[t]
\centering
\includegraphics[width=0.99\textwidth]{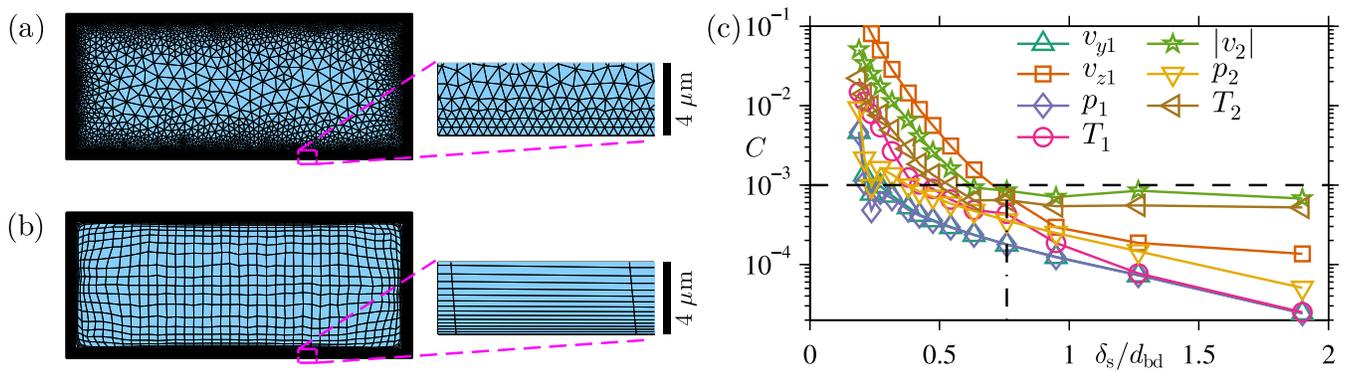}
\caption{\figlab{convergence} (Color online) (a) The used triangular mesh with a gradually increasing element size from 0.5 $\SImum$ at the boundaries to 20 $\SImum$ in the bulk. This mesh, chosen as the default mesh, contains 30246 elements. (b) Rectangular mesh with thin elongated 0.1-$\SImum$-by-10-$\SImum$ elements at the boundaries and gradually changing to nearly square 10-$\SImum$-by-10-$\SImum$ elements in the bulk. This mesh contains 3308 elements. (c) The convergence parameter $C$ of \eqref{convergenceparameter} for all first- and second-order fields versus the numerical resolution defined by $\dels/\dbd$, where $\dbd$ is the mesh-element size at the boundary. The fields are solved on triangular meshes with different boundary element sizes but all with fixed bulk element size $\dbk = 20~\SImum$ and growth rate $\alpha=1.3$, while the reference solution is calculated for $\dbd = 0.15~\SIum$, $\dbk = 2~\SIum$, and $\alpha = 1.3$. The vertical dash-dot line indicates the solution for $\dbd =0.5~\SImum$ which is chosen as the default value for the following simulations.
}
\end{figure*}%
\section{Numerical model}
The numerical scheme solves the governing equations for the acoustic field inside the two-dimensional water domain of a rectangular microchannel cross section, whereas, the vibrations in the surrounding chip material and piezo transducer are not modeled. The water domain is surrounded by immovable hard walls, and the acoustic field is excited by oscillating velocity boundary conditions, representing an oscillating nm-sized displacement of the walls. A sketch of the physical system and the numerical model is shown in \figref{chip_sketch}.

\subsection{Governing equations}
\seclab{numgov}
The governing equations are solved using the commercial software Comsol Multiphysics \cite{COMSOL44}. To achieve greater flexibility, the equations are implemented through mathematics-weak-form-PDE modules and not through the build-in modules for acoustics and fluid mechanics. In contrast to our previous work \cite{Muller2012}, the second-order equations (\ref{eq:massEq2}), (\ref{eq:momentumEq2}), and (\ref{eq:energyEq2}) are implemented in the flux density formulation, which by partial integration avoids the less accurate second-order derivatives appearing in the body-force formulation. To fix the numerical solution for the second-order mass- and momentum conservation equations, the spatial average of the second-order pressure is forced to be zero by a Lagrange multiplier.

\subsection{Boundary conditions}
\seclab{BC}
\begin{table}[t]
\caption{Acoustic impedance and thermal diffusivity for water, silicon and Pyrex glass at room temperature values from Ref. \cite{bruus2008}.\tablab{wall_param}}
\begin{ruledtabular}
\begin{tabular}{ l c  c   }
Material & Acoustic impedance  & Thermal diffusivity   \\
 & [10$^6$~kg\:m$^{-2}$\:s$^{-1}$] & [10$^{-7}$~m$^2$\:s$^{-1}$] \\ \hline
Water & $1.5$   &  $1.4$ \\
Silicon & $20$   &  $920$ \\
Pyrex glass & $17$  &  $6.3$ \\
\end{tabular}
\end{ruledtabular}
\end{table}

The first-order acoustic fields are solved in the frequency domain for a driven system, in which energy is added to the system by an oscillating velocity boundary condition and lost by thermal conduction through the walls. The walls are modeled as hard thermal conductors with infinite acoustic impedance and infinite thermal diffusivity. This approximation is reasonable given the parameter values listed in \tabref{wall_param}. In the numerical model this is implemented by zero velocity and ambient temperature at the walls
\bsub
\bal
T &= \TO, \quad \text{on all walls},\\
\vvv &= \zerovec, \quad \text{on all walls},\\
\nnn \cdot\vvvI &= \vbc(y,z)e^{-i\omega t}, \ \text{added to actuated walls,}\\
\nnn \cdot \vvvII &= -\Big\langle\frac{\rhoI}{\rhoO}(\nnn\cdot\vvvI)\!\Big\rangle, \text{added to actuated walls.}\eqlab{BCv2}
\eal
\esub
Here, $\nnn$ is the outward pointing surface normal, and \eqref{BCv2} ensures zero mass flux across the boundary.

It is not trivial how to apply the oscillating velocity boundary condition. In our model we wish to excite the horizontal half-wavelength resonance, which at the top and bottom walls leads to viscous boundary layers and the generation of streaming flow.
To avoid direct influence on this flow from the actuation, we therefore choose to actuate only the left and right walls at $y=\pm w/2$. Moreover, an optimal coupling to the half-wavelength resonance is obtained by choosing the proper symmetry of the actuation, and therefore in terms of the components $v^{{}}_{y1}$ and $v^{{}}_{z1}$, the boundary condition on $\vvvI$ becomes
\beq{boundary}
\vyI\Big(\pm \frac{w}{2},z\Big) = \vbc \ee^{-\iot},\qquad
 v^{{}}_{z1}\Big(\pm \frac{w}{2},z\Big) = 0,
\eeq
where $\vbc=\omega d$ is the amplitude of the actuation in terms of the displacement $d$, with $d=0.1$\:nm in all simulations. This velocity boundary condition is well defined and yields results consistent with experiments \cite{Muller2013}.

\subsection{Convergence analysis}
\seclab{convergence}
The weak form equations along with the boundary conditions are solved on a two-dimensional triangular mesh using the finite element method, see \figref{convergence}. The resolution of the physical field is determined by the spatial resolution of the mesh and the polynomial order of the basis functions used to represent the field in each node in the mesh. To test the validity of the numerical model we first check that the numerical solution has converged, \ie~ensuring that further refining of the mesh does not change the solution significantly.

Due to the very different length scales of the channel dimensions and the boundary layer thickness an inhomogeneous mesh is necessary, and thus there is a number of ways to refine the mesh. We used three parameters: maximum mesh-element size at the boundaries $\dbd$, maximum mesh-element size in the bulk $\dbk$, and the maximum mesh-element growth rate $\alpha$ (maximum relative size of neighboring elements). The convergence of the fields was considered through the relative convergence parameter $C(g)$ defined in Ref. \cite{Muller2012} by
\beq{convergenceparameter}
 C(g) = \sqrt{\frac{{\displaystyle \int} \big(g-g_\mathrm{ref}\big)^2\ \dm y\:\dm z}{
 {\displaystyle \int} \big(g_\mathrm{ref}\big)^2\ \dm y\:\dm z}},
\eeq
where $C(g)$ is the relative convergence of a solution $g$ with respect to a reference solution $g_\mathrm{ref}$. Convergence graphs for all fields as function of $\dbd$ are shown in \figref{convergence}(c). The mesh parameters for the reference solution are $\dbd = 0.15~\SIum$, $\dbk = 2~\SIum$, and $\alpha = 1.3$, whereas other solutions for given $\dbd$ use $\dbk = 20~\SIum$ and $\alpha = 1.3$. The basis functions for the first- and second-order velocity and temperature fields are all fourth order, while for the first- and second-order pressure they are third order. All fields exhibit good convergence, and we choose $C=10^{-3}$ as our convergence criterion in the following. The corresponding default triangular mesh has $\dbd=0.5~\SImum$, see \figref{convergence}(a). In \figref{convergence}(b) is shown a mesh with rectangular mesh elements which are nearly square in the bulk of the channel while very elongated near the walls. This mesh has been used for testing purposes as it contains approximately ten times fewer mesh elements compared to the default triangular mesh and the resulting fields all show convergence parameters below $C=10^{-3}$ with respect to the triangular reference mesh. All results have been calculated using the triangular mesh, but the square mesh provides a huge advantage regarding calculation speed and memory requirement.


\section{results}
\seclab{results}

\subsection{Resonance analysis}
\seclab{resonance}
To determine the acoustic resonance frequency $\fres$ corresponding to the horizontal half-wavelength resonance, we sweep the actuation frequency around the ideal frequency $\fO = \cs /2w$, corresponding to the half-wavelength match $\lambda /2 = w$, and calculate the acoustic energy density \eqref{Eac}, shown in \figref{resonance_curve}. The resonance frequency $\fres$ is shifted slightly with respect to the ideal frequency $\fO$ due to the viscous loss in the boundary layers. This loss also determines the width of the resonance curve and thus the Q-value of the acoustic cavity.

\begin{figure}[t]
\centering
\includegraphics[width=0.99\columnwidth]{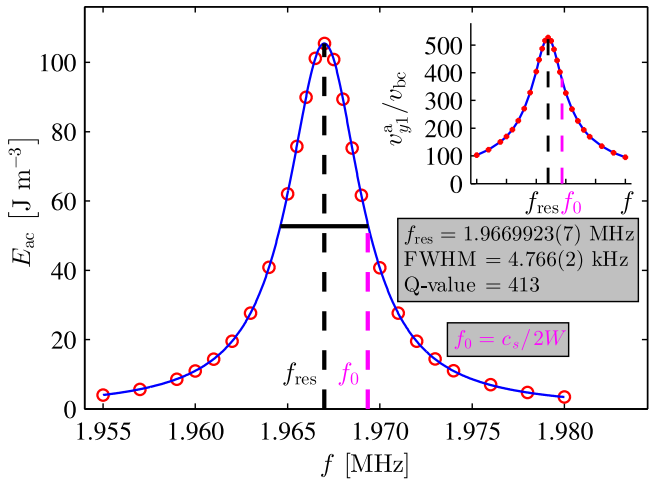}
\caption{\figlab{resonance_curve} (Color online) Graph of the acoustic energy density $\Eac$ \eqref{Eac} as function of the frequency  $f$ of the oscillating boundary condition. $\fres$ is the resonance frequency at the center of the peak, while $\fO$ is the ideal frequency corresponding to matching a half-wavelength with the channel width. The inset shows the magnitude of the resonant oscillating first-order velocity field $\vyIa$ relative to the amplitude of the oscillating velocity boundary condition $\vbc$ as function of the actuation frequency $f$.
}
\end{figure}%
%


%
\begin{figure*}[t]
\centering
\includegraphics[width=0.99\textwidth]{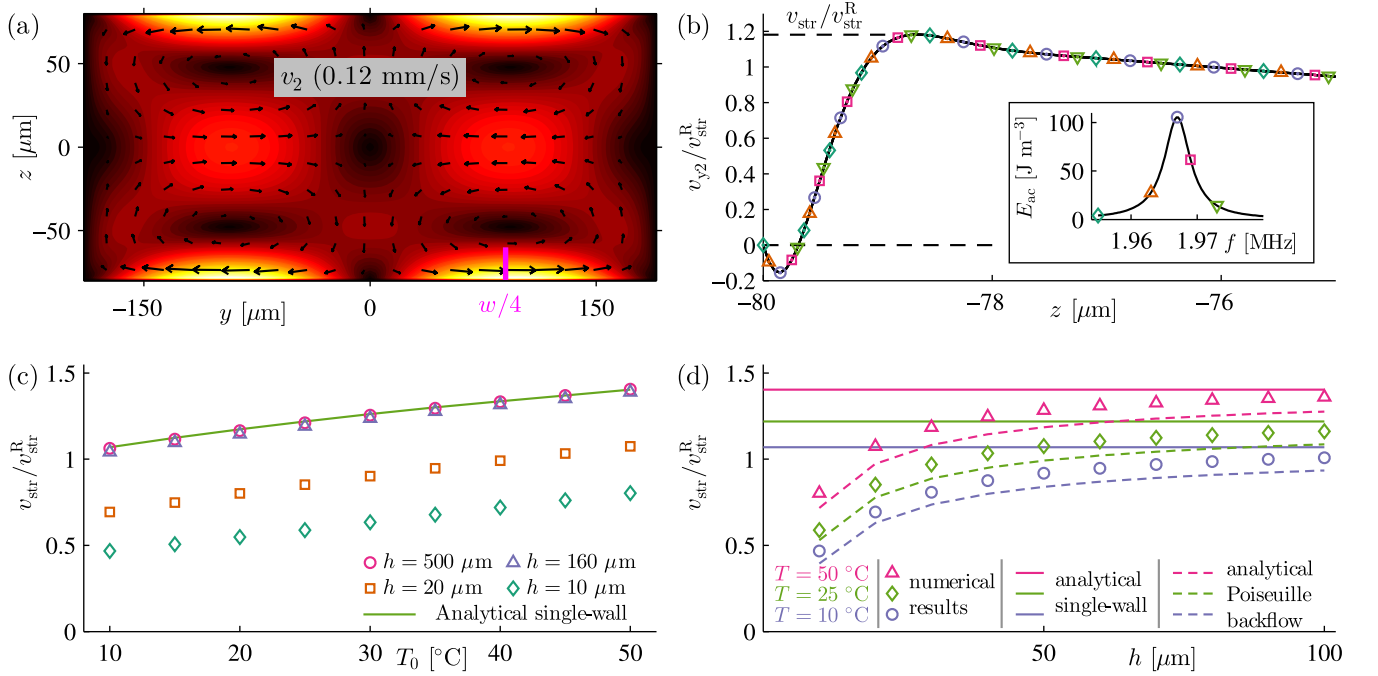}
\caption{\figlab{v2} (Color online)
(a) Time-averaged second-order fluid velocity field $\vvvII$ (vectors) and its magnitude [color plot ranging from 0 mm/s (black) to 0.12 mm/s (white)] in the vertical channel cross section calculated at $\TO=25~\SICel$ and $\fres=1.9669923$~MHz. (b) The horizontal velocity component $\vyII$ plotted along $y=w/4$ indicated by the magenta line in (a). The velocity field has been calculated for the five actuation frequencies shown in the inset resonance curve, and normalized to the analytical Rayleigh streaming magnitude $\vstrR=(3/8) (\vyIa)^2/\cs$, which is calculated based on the corresponding first-order solutions. The symbols are plotted in selected points illustrating the five numerical solutions (black lines) that coincide.
(c) Normalized streaming magnitude $\vstr/\vstrR$ (symbols) versus equilibrium temperature $\TO$, calculated for different channel heights $h$. The full curve show the analytical single-wall result by ref. \cite{Rednikov2011}. (d) Normalized streaming magnitude $\vstr/\vstrR$ (symbols) versus channel height $h$, calculated for different equilibrium temperatures $\TO$. The full curves show the analytical single-wall result by ref. \cite{Rednikov2011}, while the dashed lines show the results of a one-dimensional analytical model with a Poiseuille backflow \eqref{backflow}.
}
\end{figure*}

\subsection{Time-averaged second-order velocity}
\seclab{velocity}

The time-averaged second-order velocity field $\vvvII$ is shown in \figref{v2}(a), calculated for the default 380-$\SImum$-by-160-$\SImum$ rectangular geometry, at $\TO=25\:\SICel$, and at the resonance frequency $\fres=1.9669923$\:MHz. It exhibits the well-known pattern of four flow rolls each $\lambda/4$ wide. To investigate the magnitude of the streaming velocity, \figref{v2}(b) shows the velocity along a line perpendicular to the bottom wall at $y=w/4$. The streaming velocity field has been calculated for five different frequencies shown in the inset resonance curve.
The streaming velocities have been normalized to the classical result by Lord Rayleigh for the magnitude of the acoustic streaming $\vstrR=(3/8)(\vyIa)^2/\cs$, where $\vyIa$ is taken from the corresponding first-order solutions. The five numerical solutions (black lines) coincide completely, showing that the rescaled second-order velocity field is the same for off-resonance actuation frequencies. This is important for our further analysis, as we do not need to determine the exact resonance frequency as it changes due to variations in temperature $\TO$ and channel height $h$. The magnitude of the streaming velocity $\vstr$ is determined by the maximum value of $\vyII$ along the line $y=w/4$ as indicated in \figref{v2}(b).

In \figref{v2}(c) is shown the normalized magnitude of the streaming velocity $\vstr/\vstrR$ versus the equilibrium temperature $\TO$. The streaming velocity has been calculated for different channel heights indicated by different colors and symbols. The full line is the analytical single-wall solution by Rednikov and Sadhal \cite{Rednikov2011} for a standing acoustic wave parallel to a single planar wall. For all channel heights the streaming velocity shows an almost linear dependence with positive slope on the equilibrium temperature. The numerical results for the tall channel $h=500\:\SImum$ agree well with the analytical single-wall prediction, while for more shallow channels the steaming velocity is significantly lower. At $25~\SICel$ the streaming velocity is $19\%$ larger than the classical Rayleigh result, while for $50~\SICel$ this deviation has increased to $39\%$.

To elaborate on the dependence of the streaming velocity on the height of the channel, we plot in \figref{v2}(d) the normalized streaming velocity versus the channel height for three equilibrium temperatures. The numerical results are shown by symbols, while the analytical single-wall predictions for each temperature are shown by full lines. The numerical results for the rectangular channel deviate from the analytical single-wall prediction as the channel height is decreased. To qualitatively explain this deviation, we make a simple one-dimensional analytical model along the $z$-dimension of the rectangular channel in which we impose a boundary-driven flow. The first part of the model is a plug flow with an exponential dependence close to the wall $\vplu(z) = \vO\big\{1-\exp[-(z+h/2)/\dels]\big\}$ for $-h/2 <z<0$. This approximates the $z$-dependence of the streaming velocity field inside the viscous boundary layer, where $\vO$ corresponds to the analytical single-wall solution \cite{Rednikov2011}. As the water is pushed towards the sidewall a pressure builds up and a Poiseuille backflow is established, which by mass conservation and no-slip boundary conditions become $\vpoi(z) \approx -\vO\{6(1/4-z^2/h^2)\}$. By a first-order Taylor expansion of $\vpoi(z)$ at the wall $z=-h/2$, we can determine the maximum $\vstr$ of $\vplu(z) + \vpoi(z)$ near the wall to first order in $\dels/h$,
\beq{backflow}
\vstr \approx \vO\Bigg\{1-6\frac{\dels}{h}\bigg[1+\ln\bigg(\frac{h}{6\dels}\bigg)\bigg]\Bigg\}.
\eeq
This provides an estimate for the magnitude of the acoustic streaming shown by dashed lines in \figref{v2}(d), with the viscous boundary layer thickness \eqref{dels} calculated for each of the three temperatures. This simple one-dimensional analytical model captures the trend of the numerical data well, though overall it predicts slightly lower streaming amplitudes. The deviation from the numerical data is ascribed primarily to the monotonic approximation $\vplu(z)$ of the $z$-dependence of the velocity inside the viscous boundary layer. The full $z$-dependence of the streaming velocity inside the viscous boundary layer is non-monotonic and overshoots slightly before leveling. This can be seen in \figref{v2}(b), and thus the maximum velocity occurs at this overshoot and is consequently slightly larger than predicted by the approximate analytical model. For channel heights below 10~$\SImum$ the assumptions of a boundary driven plug flow with a superimposed Poiseuille backflow begins to collapse as the height of the channel becomes comparable to the boundary layer thickness, and a more elaborate analytical calculation of the streaming velocity field is necessary \cite{Hamilton2003}.


\subsection{Time-averaged second-order temperature}
\seclab{temperature}
In \figref{T2}(a) is shown the time-averaged second-order temperature field $\TII$ calculated for the default 380-$\SImum$-by-160-$\SImum$ geometry at the resonance frequency. In \figref{T2}(b-c) are shown line plots of $\TII$ along the horizontal and vertical dashed lines in \figref{T2}(a). $\TII$ has a saddle point in the center of the channel $(y=0,z=0)$, two global maxima on the horizontal centerline $z=0$, and a wide plateau on the vertical center line $y=0$. The temperature field is forced to be zero at all boundaries due to the boundary condition of infinite heat conduction. The gradient of $\TII$ along line C indicates a decline in heat generation inside the boundary layer going from the center towards the left and right walls. The global maxima in the bulk result from heat generation in the bulk as discussed in \secref{discussion}.
\begin{figure}
\centering
\includegraphics[width=0.99\columnwidth]{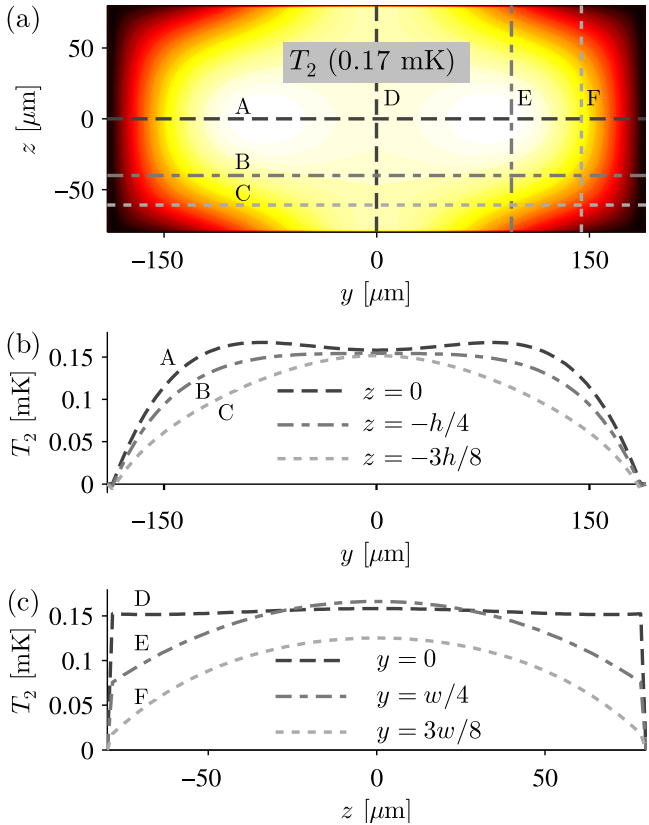}
\caption{\figlab{T2} (Color online) Time-averaged second-order temperature $\TII$ calculated for the default 380-$\SImum$-by-160-$\SImum$ geometry, actuation frequency $\fres = 1.9669923$\:MHz, and equilibrium temperature $\TO=25\:\SICel$. (a) Color plot (black 0\:mK to white 0.17\:mK) of $\TII$ in the channel cross section. (b-c) Line plots of $\TII$ along the horizontal and vertical dashed lines in (a), respectively.}
\end{figure}

\section{Discussion}
\seclab{discussion}
\begin{figure}
\centering
\includegraphics[width=0.99\columnwidth]{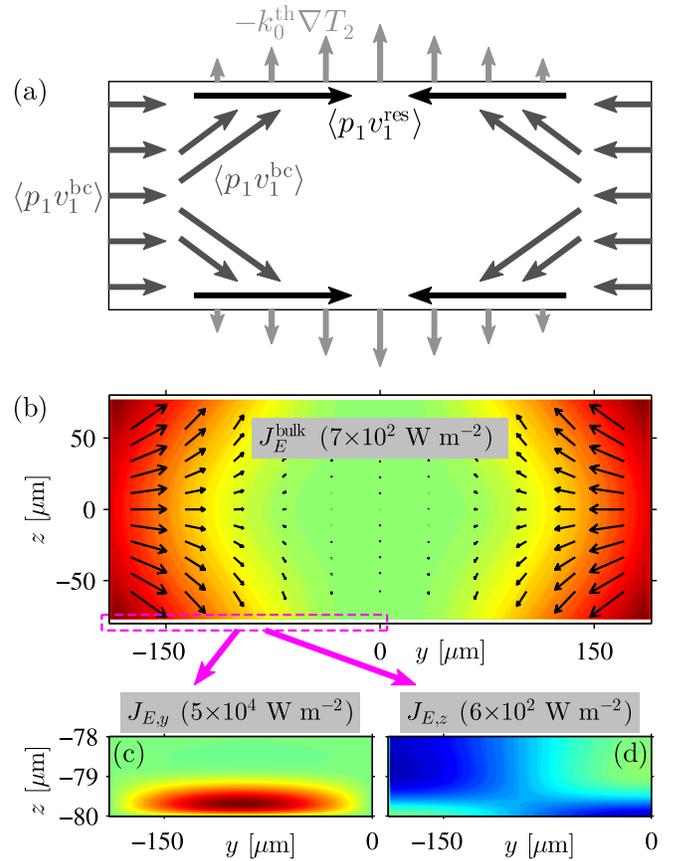}
\caption{\figlab{JE} (Color online) Time-averaged heat current densities in the channel cross section. (a) Sketch of the heat currents (arrows) with indication of the responsible terms in the time-averaged energy conservation equation \eqref{energyEq2all}. (b) Heat current density (arrows) and its magnitude [color plot ranging from zero (light green) to $7\times10^2$~W~m$^{-2}$ (dark red)] in the bulk of the channel. The strong currents inside the boundary layers are not shown. (c) Magnitude of the $y$-component of the heat current density [color plot ranging from zero (light green) to $5\times10^4$~W~m$^{-2}$ (dark red)] inside the boundary layer at the bottom wall. (d) Magnitude of the $z$-component of the heat current density [color plot ranging from $-6\times10^2$~W~m$^{-2}$ (dark blue) to zero (light green)] inside the boundary layer at the bottom wall}
\end{figure}
\begin{figure}
\centering
\includegraphics[width=0.99\columnwidth]{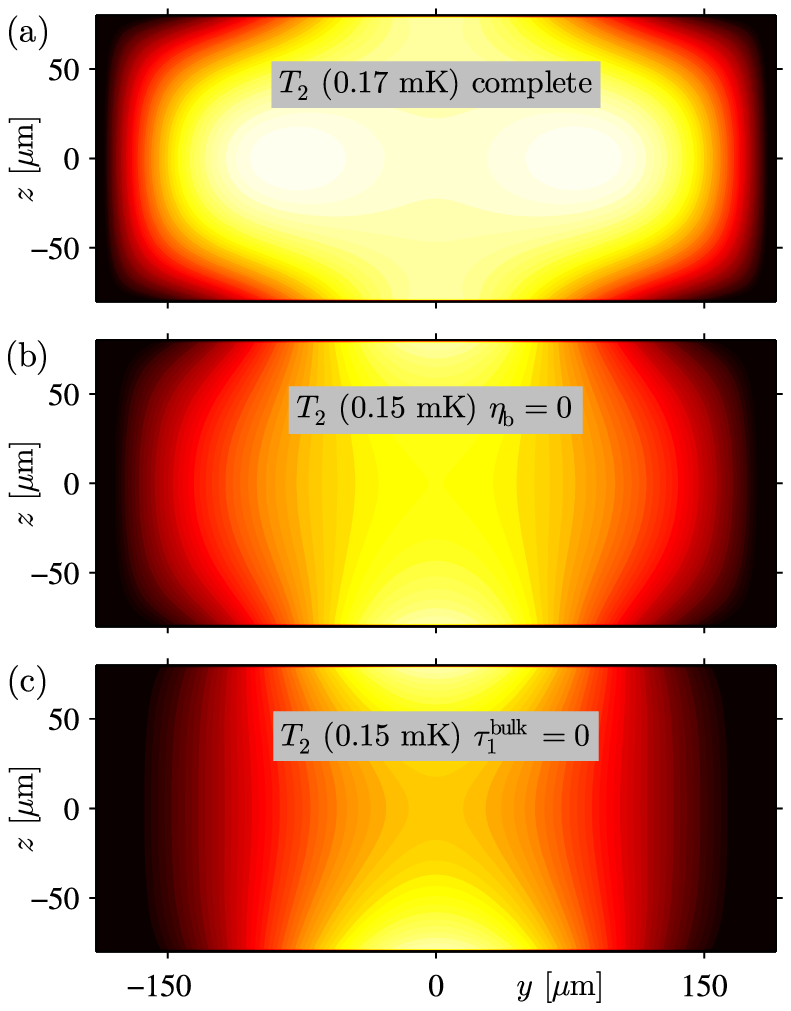}
\caption{\figlab{T2_arti} (Color online) Color plot of the time-averaged second-order temperature $\TII$ [from zero (black) to maximum (white)] in three cases. (a) $\TII$ calculated from the complete governing equations, identical to \figref{T2}(a). (b) $\TII$ calculated without bulk viscosity $\etab=0$. (c) $\TII$ calculated with zero viscous stress $\bm{\tau}=\bm{0}$ in the bulk defined by $\vert y\vert < (w/2-4~\SImum)$ and $\vert z\vert < (h/2-4~\SImum)$.}
\end{figure}
In \figref{JE}(a) we provide an overview of the energy transport and dissipation in the system by showing a sketch of the energy currents in the channel cross section. To explain the convection of energy, we consider the first-order velocity to be composed of a weak non-resonant part $\vvvIbc$, which fulfills the oscillating velocity boundary conditions, and a strong resonant part $\vvvIres$, which has zero amplitude at all walls \cite{Bruus2012}. In \figref{JE}(b) is shown the total energy current density, given by all the terms inside the divergence in \eqref{energyEq2}, in the bulk of the channel, thus not including the thin boundary layers at the top and bottom walls. The plot shows how mechanical energy is entering the system at the left and right walls, due to the oscillating boundary condition, and is convected towards the top and bottom walls. This transport is dominated by the term $\avr{\pI\vvvI}$ in \eqref{energyEq2}, particularly the non-resonant part $\avr{\pI\vvvIbc}$, since $\vvvIres$ is out of phase with $\pI$ in the bulk. The $y$- and $z$-component of the energy current density inside the boundary layer at the bottom wall is shown in \figref{JE}(c-d). The transport parallel to the wall, \figref{JE} (c), results from $\avr{\pI\vvvIres}$, which is large, since $\vvvIres$ is phase shifted inside the boundary layer. The transport perpendicular to the wall, \figref{JE}(d), results predominantly from the thermal diffusion term $-\kthO\nablabf\TII$.

To rationalize the amplitudes of the fields we estimate the order of magnitude of the energy transport and dissipation in the system. The incoming energy current density from the oscillating velocity boundary condition at the left and right walls is given by the time-averaged product of the local pressure and velocity $\avr{\pI \vyIbc }$. Multiplying this by the area $2h\ell$, we obtain the magnitude of the incoming power $\Pin \sim 2h\ell\frac{1}{2}\pIa \vbc$. Here, the factor $\frac{1}{2}$ enters from time averaging, $\ell$ is the channel length, and the superscript ``$\mathrm{a}$'' denotes the amplitude of the resonant field. From the inviscid part of the first-order momentum conservation \eqref{momentumEq1}, we estimate the magnitude $\pIa \sim \rhoO \cs \vyIa$ and therefore obtain $\Pin \sim h\ell \rhoO \cs \vyIa \vbc$

The dissipation of mechanical energy happens primarily in the viscous boundary layers of thickness $\dels$ due to the work done by the viscous stress force density $(\nablabf \cdot \taubfI)$ with power density $\avr{(\nablabf \cdot \taubfI)\cdot\vvvI}$. As the gradient of $\vvvI$ perpendicular to the wall inside the boundary layer is large, the dominant term is $\avr{\etaO\frac{\partial^2 \vyI}{\partial z^2}\:\vyI} \sim \etaO\frac{1}{4}(\vyIa)^2/\delssqr$, where two factors of $\frac{1}{2}$ enters from spatial and time averaging. The total power dissipation is given by the product of the power density and the volume of the boundary layers $\Ps \sim 2\dels w\ell\etaO\frac{1}{4}(\vyIa)^2/\delssqr$.

In steady state $\Ps$ equals $\Pin$, from which we find the magnitude $\vyIa$ of the resonant field in terms of $\vbc$ to be $\vyIa \sim \frac{2}{\pi} \frac{h}{\dels}\frac{\lambda}{w} \vbc \sim 500\vbc$ for our system, which is in good agreement with the numerical result for $\vyIa/\vbc$ plotted in the inset of \figref{resonance_curve}.

To rationalize the magnitude of the second-order temperature shift, we consider the diffusive energy transport through the top and bottom walls. The diffusive energy current density is $-\kthO\nablabf\TII$, and as heat diffuses to the perfectly conducting walls on a length scale of $\delt$, the outgoing power is $\Pout \sim 2w\ell\kthO(\tfrac{1}{2}\TIIa/\delt)$. Here, the spatial average of $\TII$ just outside the thermal boundary layers along the top and bottom walls has been approximated to $\tfrac{1}{2}\TIIa$. In steady state $\Pout$ equals $\Pin$ and the magnitude of the second-order temperature becomes
$\TIIa \sim  \frac{2}{\pi^2}\frac{h^2}{\delt\dels}(\frac{\lambda}{w})^2 \frac{1}{\cp}(\vbc)^2 \sim 0.13$~mK, which is comparable to the numerical result in \figref{T2}.

From the simplified picture of strong heat generation inside the boundary layers it may seem odd that the second-order temperature field in \figref{T2} has two global maxima in the bulk of the channel. This effect is due to the absorption in the bulk of the channel originating from the non-zero divergence of the stress force term $\avr{\vvvI\cdot\taubfI}$ in \eqref{energyEq2} as shown in \figref{T2_arti}. In \figref{T2_arti}(a) is shown the complete second-order temperature field $\TII$. \figref{T2_arti}(b) shows an artificial temperature field calculated without bulk viscosity, $\etab = 0$. No maxima appears in the bulk, and the temperature field looks more as expected from the simplified view of heat generation in the boundary layers. However, there is still a small heat generation in the bulk of the channel from the shear viscosity. In \figref{T2_arti}(c) this heat generation is suppressed by setting $\taubfI = \zerovec$ in the bulk more than 4~$\SImum$ from the walls, while maintaining the full $\taubfI$ in the boundary layers. The resulting plot of $\TII$ shows how heat is generated in boundary layers near the top and bottom walls and mainly conducted out of these, while some of the heat is conducted into the bulk and out through the left and right walls. The bulk viscosity $\etab$ is often neglected when working at frequencies around 2~MHz because of its small contribution to the total dissipation, and the subsequent negligible influence on the resonance curve and the streaming velocity field. However, \figref{T2_arti} clearly shows that the bulk absorption is important for the spatial structure of the time-averaged temperature field.

In \secref{IAPWS} we stated that the changes in the dynamic viscosity due to its temperature and density dependence are $0.33\%$ and $0.37\%$, respectively, for the amplitudes of the acoustic oscillation used in this paper. It might seem surprising that, firstly, such a small perturbation of the viscosity can increase the magnitude of the streaming by $19\%$ at $25\SICel$ as stated in \secref{velocity} ($39\%$ at $50\SICel$), and secondly, the numerical results are in very good agreement with the analytical expression from Ref. \cite{Rednikov2011}, which does not include the density dependence of the dynamic viscosity of similar magnitude as the temperature dependence. The explanation lies within the spatial structure of the fields. From the time-averaged momentum equation (\ref{eq:momentumEq2both}), we see that the divergence of the stress tensor leads to a term containing the gradient of the viscosity perturbation $\big\langle\!\nablabf\etaI\cdot\big[\nablabf\vvvI\! + (\nablabf \vvvI)^\mathrm{T}\big]\big\rangle$. Here, $\etaI$ is proportional to $\TI$ and $\rhoI$, and since $\TI$ changes on the small length scale of the boundary layer, whereas $\rhoI$ only changes on the long length scales of the channel width, we get $\nablabf\etaI = \nablabf\eta^{(T)}_1 + \nablabf\eta^{(\rho)}_1 \sim \eta^{(T)}_1/\delt + \eta^{(\rho)}_1/w \approx \eta^{(T)}_1/\delt$, where the superscripts refer to the contribution from either the temperature or the density perturbation. Consequently, with respect to the acoustic streaming, the temperature dependence of the dynamic viscosity is much more important than the density dependence.

The significant increase of the acoustic streaming magnitude, due to the temperature-induced viscosity perturbation, influences the interplay between radiation forces and drag forces on suspended particles \cite{Muller2012,Barnkob2012a}. The steady temperature rise of less than 1~mK has on the other hand negligible influence on acoustic handling of biological samples, however, other applications of acoustofluidics, such as thermoacoustic engines, rely on the steady energy currents for pumping heat from a low-temperature source to a high-temperature sink, or inversely, for generating acoustic power from the heat flow between a high-temperature source and a low-temperature sink \cite{Rott1975,Swift1988}.


\section{Conclusion}
\seclab{conclusion}
In this work, we have presented a full numerical study of the acoustic streaming in the cross section of a long straight microchannel including the temperature and density dependence of the fluid viscosity and thermal conductivity. The temperature dependence of the streaming amplitude in the case of a deep microchannel agreed well with the analytical single-wall result from 2011 by Rednikov and Sadhal \cite{Rednikov2011}, whereas significant deviations were found for shallow channels. This strong dependence of the streaming amplitude on the channel height was explained qualitatively with a simple one-dimensional backflow model. Furthermore, we showed that a meaningful comparison of solutions at different temperatures and off-resonance frequencies could be performed by normalizing the second-order velocity field to the square of the first-order velocity amplitude.

We have also solved the time-averaged second-order energy conservation equation numerically and calculated the steady temperature rise in the channel, as well as analyzed the energy transport in the system. For acoustophoretic devices, the temperature rise is less than 1~mK and has no consequences for neither operation conditions nor biological samples. However, in other application such as thermoacoustic engines, the energy transport is important.

Finally, we have provided polynomial fits in the temperature range from $10~\SICel$ to $50~\SICel$ of the thermodynamic properties and transport properties of water at ambient atmospheric pressure based on data from IAPWS which covers a much wider range of temperatures and pressures. This allows for easy implementation of the official parameter values for the properties of water in other models working under the same temperature and pressure conditions.

With the inclusion of the local perturbation in viscosity and thermal conductivity, due to their temperature and density dependence, we have solved the complete time-averaged second-order acoustic equations for a Newtonian fluid enclosed by vibrating walls, with the one exception of the unknown density dependence of the bulk viscosity. To further progress the numerical analysis of microchannel acoustic streaming, one should improve the modeling of the vibration of the walls, preferably including the elastic waves in the surrounding solid material. In the present model the acoustic streaming velocity field depends strongly on the choice of actuation conditions on the walls.

\acknowledgments

We thank Prof.\ em.\ Dr.-Ing.\ Wolfgang Wagner, Ruhr-Universit\"{a}t Bochum, for providing us with the software FLUIDCAL, Version Water (IAPWS-95), for calculating the thermodynamic properties of water.
This work was supported by the Danish Council for Independent Research, Technology, and Production Sciences (grant no. 11-107021).

\begin{figure*}[t!]
\centering
\includegraphics[width=0.99\textwidth]{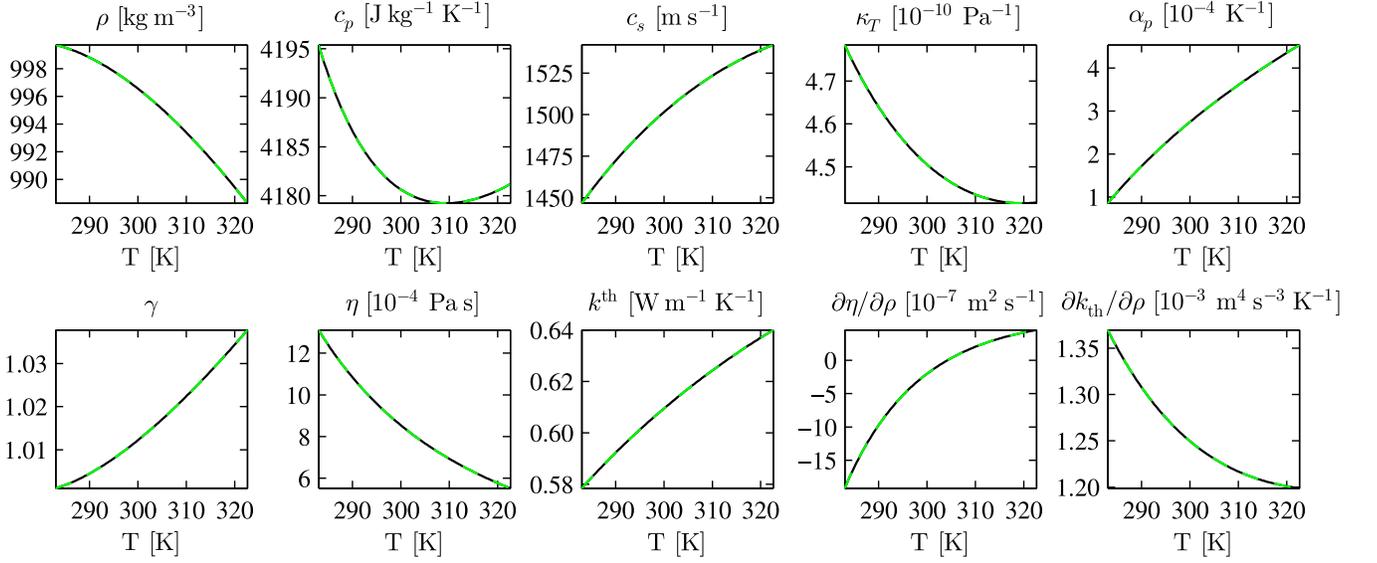}
\caption{\figlab{fits} (Color online) Fifth-order polynomial fits (dashed light lines) to IAPWS data \cite{Wagner2002, Huber2009, Huber2012} (full black lines) of ten thermodynamic and transport parameters of water in the temperature range from 283~K ($10~\SICel$) to 323~K ($50~\SICel$).}
\end{figure*}

\begin{ruledtabular}
\begin{table*}[t!]
\caption{\label{tab:fit_results}
List of the coefficients $C_i$ of the fifth-order polynomial fits to the IAPWS data for water \cite{Wagner2002, Huber2009, Huber2012}, shown in \figref{fits}. The temperature dependence of each of the ten termodynamic or transport parameters $g$ with SI unit $[g]$ is fitted in the temperature range from from 283~K ($10~\SICel$) to 323~K ($50~\SICel$) by a polynomial of the form $g(\tilde{T})/[g] = C_0 + C_1\tilde{T} + C_2\tilde{T}^2+ C_3\tilde{T}^3 + C_4\tilde{T}^4 + C_5\tilde{T}^5$, with $\tilde{T} = T/(1~\SIK)$.  MD and AD denote the maximum and average relative deviation, respectively, of the fit from the data. Text files with the polynomial coefficients are provided in the Supplementary Material~[37].}
\begin{tabular}{c@{}r@{$\!\!$}l@{}r@{$\!\!$}l@{}r@{$\!\!$}l@{}r@{$\!\!$}l@{}r@{$\!\!$}l}
      &     & \hspace*{10mm} $\rho$        &      & \hspace*{10mm} $c_p$          &       & \hspace*{10mm} $c^{{}}_s$      &       & \hspace*{10mm} $\kappa^{{}}_T$   &       & \hspace*{10mm} \text{$\alpha^{{}}_p$} \\ \hline
$C_0$ &$-8.$&$8657129122\!\times\!10^{+3}$ &$  2.$&$9011416353\!\times\! 10^{+5}$ & $ -3.$&$1420310908\!\times\! 10^{+4} $ & $  1.$&$4812304299\!\times\! 10^{-7} $   & $ -6.$&$2142868017\!\times\! 10^{-1} $ \rule{0mm}{4mm} \\
$C_1$ &$ 1.$&$4835428749\!\times\!10^{+2}$ &$ -4.$&$4547002259\!\times\! 10^{+3}$ & $  4.$&$4207693183\!\times\! 10^{+2} $ & $ -2.$&$2647220701\!\times\! 10^{-9} $   & $  9.$&$6088951712\!\times\! 10^{-3} $ \\
$C_2$ &$-8.$&$9906362105\!\times\!10^{-1}$ &$  2.$&$7826280608\!\times\! 10^{+1}$ & $ -2.$&$4555531077\!\times\! 10^{+0} $ & $  1.$&$3993170593\!\times\! 10^{-11} $  & $ -5.$&$9809671543\!\times\! 10^{-5} $ \\
$C_3$ &$ 2.$&$7498750299\!\times\!10^{-3}$ &$ -8.$&$7096216461\!\times\! 10^{-2}$ & $  7.$&$0411177557\!\times\! 10^{-3} $ & $ -4.$&$3493756936\!\times\! 10^{-14} $  & $  1.$&$8708110371\!\times\! 10^{-7} $ \\
$C_4$ &$-4.$&$2441616455\!\times\!10^{-6}$ &$  1.$&$3656418313\!\times\! 10^{-4}$ & $ -1.$&$0353739103\!\times\! 10^{-5} $ & $  6.$&$7914632960\!\times\! 10^{-17} $  & $ -2.$&$9366903765\!\times\! 10^{-10} $ \\
$C_5$ &$ 2.$&$6348942330\!\times\!10^{-9}$ &$ -8.$&$5786317870\!\times\! 10^{-8}$ & $  6.$&$1949139692\!\times\! 10^{-9} $ & $ -4.$&$2558992023\!\times\! 10^{-20} $  & $  1.$&$8495258230\!\times\! 10^{-13} $ \\
MD    &$ 0.$&$1\!\times\! 10^{-6}        $ &$  1.$&$2\!\times\! 10^{-6}         $ & $  0.$&$2\!\times\! 10^{-6}          $ & $  5.$&$3\!\times\! 10^{-6} $            & $  1.$&$244\!\times\! 10^{-4} $ \\
AD    &$ 0.$&$0\!\times\! 10^{-6}        $ &$  0.$&$3\!\times\! 10^{-6}         $ & $  0.$&$1\!\times\! 10^{-6}          $ & $  1.$&$4\!\times\! 10^{-6} $            & $  0.$&$119\!\times\! 10^{-4} $ \\ \hline\hline
\rule{0mm}{3.5mm}
      &     &\hspace*{10mm} $\gamma$       &      & \hspace*{10mm} $\eta$         &       & \hspace*{10mm}$k^\mathrm{th}$  &&\hspace*{8mm}$\partial \eta/\partial\rho$&       &\hspace*{5mm}\text{$\partial k^{{}}_\mathrm{th}/\partial\rho$}  \\ \hline
$C_0$ &$ 8.$&$6095341563\!\times\!10^{+1} $&$  3.$&$8568288635\!\times\! 10^{+0}$ & $ -4.$&$5378052364\!\times\! 10^{+1} $ & $ -3.$&$7043919290\!\times\! 10^{-2} $   & $  3.$&$3029688424\!\times\! 10^{-1} $ \rule{0mm}{4mm} \\
$C_1$ &$-1.$&$3297233662\!\times\!10^{+0} $&$ -6.$&$0269041999\!\times\! 10^{-2}$ & $  7.$&$0158464759\!\times\! 10^{-1} $ & $  5.$&$8520864113\!\times\! 10^{-4} $   & $ -4.$&$6998644326\!\times\! 10^{-3} $ \\
$C_2$ &$ 8.$&$3329451753\!\times\!10^{-3} $&$  3.$&$7823493660\!\times\! 10^{-4}$ & $ -4.$&$3372914723\!\times\! 10^{-3} $ & $ -3.$&$7051700834\!\times\! 10^{-6} $   & $  2.$&$7065291331\!\times\! 10^{-5} $ \\
$C_3$ &$-2.$&$6194143444\!\times\!10^{-5} $&$ -1.$&$1905791749\!\times\! 10^{-6}$ & $  1.$&$3522238266\!\times\! 10^{-5} $ & $  1.$&$1749472916\!\times\! 10^{-8} $   & $ -7.$&$8584696013\!\times\! 10^{-8} $ \\
$C_4$ &$ 4.$&$1304638557\!\times\!10^{-8} $&$  1.$&$8785807901\!\times\! 10^{-9}$ & $ -2.$&$1181858883\!\times\! 10^{-8} $ & $ -1.$&$8657255651\!\times\! 10^{-11} $  & $  1.$&$1504432763\!\times\! 10^{-10} $ \\
$C_5$ &$-2.$&$6113941825\!\times\!10^{-11}$&$ -1.$&$1881783857\!\times\!10^{-12}$ & $  1.$&$3309059289\!\times\! 10^{-11}$ & $  1.$&$1866022487\!\times\! 10^{-14} $  & $ -6.$&$7916212133\!\times\! 10^{-14} $ \\
MD    &$ 1.$&$6\!\times\! 10^{-6}         $&$  1.$&$379\!\times\! 10^{-4}       $ & $  1.$&$0\!\times\! 10^{-6}          $ & $ 15.$&$459\!\times\! 10^{-4} $          & $  0.$&$8\!\times\! 10^{-6} $ \\
AD    &$ 0.$&$4\!\times\! 10^{-6}         $&$  0.$&$227\!\times\! 10^{-4}       $ & $  0.$&$3\!\times\! 10^{-6}          $ & $  3.$&$851\!\times\! 10^{-4} $          & $  0.$&$1\!\times\! 10^{-6} $ \\
\end{tabular}
\end{table*}
\end{ruledtabular}

\appendix
\section{IAPWS formulation}
\seclab{appendix}

To ease the use of the official IAPWS values for the thermodynamic and transport properties of water in our numerical analysis, we fit polynomials in temperature to the data. The precise fitting procedure and its validation are described in the following.

The data for the thermodynamic properties is obtained from an Excel implementation \cite{fluidcalIAPWS95} of the
IAPWS Formulation 1995 \cite{Wagner2002}, in which the equation of state for water is fitted using a function with 56 parameters covering the range $T^{{}}_\mathrm{melt}\leq T\leq 1273.15~\SIK$ and $p\leq 1000$~MPa.
The shear viscosity is taken from the IAPWS Formulation 2008 \cite{Huber2009}, and the thermal conductivity is taken from the IAPWS Formulation 2011 \cite{Huber2012}, for which we have implemented the expressions stated in the papers to extract data values in the temperature and density range of interest to us. The data for the density derivatives of the viscosity and the thermal conductivity has been obtained using a central difference $\partial\eta/\partial \rho\approx[\eta(T,\rho+\dd\rho)-\eta(T,\rho-\dd\rho)]/(2\dd\rho)$, with $\dd\rho=0.001$~kg\:m$^{-3}$. The bulk viscosity is taken from Holmes, Parker, and Povey \cite{Holmes2011}, who extended the work by Dukhin and Goetz \cite{Dukhin2009}. The former paper provides  a third-order polynomial fit in temperature to $\etaB$, thus rendering further fitting superfluous.

For each parameter, we extract 400 data points equally spaced in the temperature range from 283~K to 323~K at ambient pressure $p=0.101325$~MPa. From these 400 values, we only use every fourth point for the fitting of a fifth-order polynomial, while the remaining 300 data points are used for calculating the deviation of the fit from the data. The order of the polynomial has been chosen as a tradeoff between low deviation between fit and data, achieved at high polynomial order, and low uncertainty in the polynomial coefficients, achieved at low polynomial order. The fitted polynomial coefficients are then truncated to a finite precision of 11 significant digits. The number of significant digits has been chosen such that the finite precision of the fitting coefficients does not result in larger deviations between fit and data. We then calculated the relative deviation of the fit with respect to each of the 400 data points and derive the maximum relative deviation (MD) and the average relative deviation (AD). The derivative of the shear viscosity with respect to the density is a special case since it crosses zero in the temperature interval, and thus the deviation between the fit and the data points are normalized to the mean of all the data points instead of the local data point. The data points and the polynomial fits for all fitted parameters are shown in \figref{fits}, and the polynomial coefficients and the maximum and average relative deviation between fit and data are shown in \tabref{fit_results}. The fitted coefficients are provided in the Supplementary Material \footnote{See Supplemental Material at [URL] for text files with the fitted polynomial coefficients for the temperature dependence of the thermodynamic and transport parameters, both in a general format \texttt{coefficients\_general\_format.txt} for copy-paste use, and a format \texttt{coefficients\_comsol\_format.txt for direct import into Comsol Multiphysics}}.

%
%


%

\end{document}